\documentstyle[aps,floats,epsfig,axodraw,amsmath,graphicx,subfigure]{revtex}
\input epsf

\begin{document}
%
\twocolumn[\hsize\textwidth\columnwidth\hsize\csname
@twocolumnfalse\endcsname
%
\author{Michael Doran\footnotemark[1] and  J\"org J\"ackel\footnotemark[2]}
\title{Loop corrections to scalar quintessence potentials}

\address{Institut f\"ur Theoretische Physik der Universit\"at Heidelberg\\
  Philosophenweg 16, D-69120 Heidelberg, Germany}
 
\maketitle
 
\pagenumbering{arabic}     
\begin{abstract}
  The stability of scalar quintessence potentials under quantum
  fluctuations is investigated for both  uncoupled models and models
  with a coupling to fermions. We find that uncoupled models are usually
  stable in the late universe. However, a coupling to fermions is
  severely restricted. We check whether a graviton induced
  \mbox{fermion-quintessence} coupling is compatible with this restriction.
\end{abstract}

%
 ]

\footnotetext[1]{\tt doran@thphys.uni-heidelberg.de}
\footnotetext[2]{\tt jaeckel@thphys.uni-heidelberg.de}
\renewcommand{\thefootnote}{\arabic{footnote}}      

\newcommand{\omd}{\Omega _{\rm d}}
\newcommand{\sa}{\sigma _8}
\newcommand{\omdsf}{\bar{\Omega} _{\rm d} ^{\rm sf}}
\newcommand{\omdn}{\Omega _{\rm d}^0}
\newcommand{\omdeq}{\sqrt{1-\bar{\Omega}_{\rm d} (\aeq)}}
\newcommand{\omdroot}{\sqrt{1-\bar{\Omega}_{\rm d} (a)}}
\newcommand{\aeq}{a_{\rm eq}}
\newcommand{\adec}{a_{\rm dec}}
\newcommand{\atr}{a_{\rm tr}}
\newcommand{\wda}{w_{\rm d}}
\newcommand{\wdan}{w_{\rm d}^0}
\newcommand{\kmax}{k_{\rm max}}
\newcommand{\keq}{k_{\rm eq}}
\newcommand{\sla}[1]{\slash\!\!\!#1}
\newcommand{\slad}[1]{\slash\!\!\!\!#1}
\newcommand{\lferm}{\Lambda_{\rm ferm}}
\newcommand{\pmp} {\Phi} 
\newcommand{\pmpc}{\Phi_{\rm cl}} 
\newcommand{\gev}{\textrm{Gev}}
\newcommand{\eloop}{\textrm{1-loop}}
\newcommand{\vv}[2]{V_{\textrm{#1}}^{\textsc{#2}}}
\newcommand{\ww}[2]{W_{\textrm{#1}}^{\textsc{#2}}}
\newcommand{\fpi}{(2\pi)^{-4}}
\newcommand{\tpi}{(2\pi)^{-2}}
\newcommand{\mf}{m_{\rm f}(\Phi_{\rm cl})}
\newcommand{\mfn}{m_{\rm f}^0}
\newcommand{\mfnsq}{\left[\mfn\right]^2}
\newcommand{\mfsq}{\left[\mf\right]^2}
\newcommand{\mplank}{\textrm{M}_{\textrm{P}}}

\def\frac#1#2{\mathinner{#1\over#2}}

\section{Introduction}\label{introduction}
Observations indicate that dark energy constitutes a substantial
fraction of our Universe
\cite{Riess:1998cb,Perlmutter:1999np,Netterfield:2001yq,Lee:2001yp,twodf}.
The range of possible candidates
includes a cosmological constant and -- more flexibly -- some
form of dark energy with a time dependent equation of state, called
quintessence \cite{Caldwell:1998ii}.  Commonly, realizations of  quintessence scenarios
feature a light scalar field \cite{Peebles:1988ek,Ratra:1988rm,Wetterich:1988fm}.

The  evolution of the scalar field is usually treated at the
classical level. However, quantum fluctuations may alter the
classical quintessence potential. In this article, we will
investigate one-loop contributions  to the effective potential from both 
quintessence and fermion fluctuations.
We will show that in the late universe, quintessence fluctuations
are harmless for most of the potentials used in the literature.
For inverse power laws and SUGRA inspired models, this has already
been demonstrated in \cite{Brax:2000yv}. That the smallness of the quintessence 
mass   needs to be protected by some symmetry has
been pointed out in \cite{Kolda:1999wq,Peccei:2000rz}.

In contrast with the rather harmless quintessence field fluctuations,
fermion  fluctuations severely restrict the
magnitude of a possible coupling of quintessence to fermionic dark
matter, as we will show.

In Euclidean conventions, the action we use for the quintessence
field $\Phi$ and a fermionic species $\Psi$ to which it may couple
\cite{Wetterich:1988fk,Amendola:2000er,Bean:2001zm} is
\begin{multline}\label{equ::action}
S = \int d^4x\  \sqrt{g}\Bigg[\mplank^2R+\frac{1}{2} \partial_{\mu} \Phi(x) \partial^{\mu} \Phi(x) + V(\Phi(x)) \\
+ \bar\Psi(x)\left[i\ \slad{\nabla} + \gamma^5 m_{\rm f}(\Phi)  \right ]  \Psi(x)\Bigg],
\end{multline}
with $m_{\rm f}(\Phi)$ as a $\Phi$ dependent fermion mass. This $\Phi$
dependence (if existent in a model) determines the coupling
of the quintessence field to the fermions.
As long as one is not interested in quantum gravitational
effects, one may set  $\sqrt{g}=1$, $R=0$
and replace $\slad{\nabla}\rightarrow\sla{\partial}$
in  the  action \eqref{equ::action}.

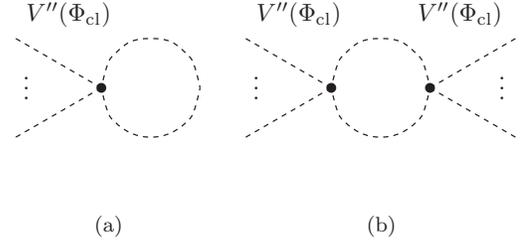
\begin{figure}[!t]
\begin{center}

\subfigure[]{\scalebox{0.93}[0.93]{
\begin{picture}(90,70)\label{fig::pure1}
\SetOffset(-100,-25)
\Text(110,90)[l]{$V''(\pmpc)$}
\Vertex(140,60){2}
\DashLine(140,60)(105.36,40){2}
\DashLine(140,60)(105.36,80){2}
\Text(110,63)[]{$\vdots$}
\DashCurve{(140,60)(142.7,50)(145.86,45.86)(150,42.7)(160,40)(170,42.7)(174.14,45.86)(177.3,50)(180,60)}{2}
\DashCurve{(140,60)(142.7,70)(145.86,74.14)(150,77.7)(160,80)(170,77.7)(174.12,74.14)(177.3,70)(180,60)}{2}
\end{picture}}
}
\subfigure[]{\scalebox{0.93}[0.93]{
\begin{picture}(125,70) \label{fig::pure2}
\SetOffset(-100,-25)
\Text(110,90)[l]{$V''(\pmpc)$}
\Vertex(140,60){2}
\DashLine(140,60)(105.36,40){2}
\DashLine(140,60)(105.36,80){2}
\Text(110,63)[]{$\vdots$}
\DashCurve{(140,60)(142.7,50)(145.86,45.86)(150,42.7)(160,40)(170,42.7)(174.14,45.86)(177.3,50)(180,60)}{2}
\DashCurve{(140,60)(142.7,70)(145.86,74.14)(150,77.7)(160,80)(170,77.7)(174.12,74.14)(177.3,70)(180,60)}{2}
\Text(210,90)[r]{$V''(\pmpc)$}
\Vertex(180,60){2}
\DashLine(180,60)(214.64,40){2}
\DashLine(180,60)(214.64,80){2}
\Text(210,63)[]{$\vdots$}
\end{picture}}
}
\caption{Pure quintessence fluctuations (depicted as dashed lines). The loop of the fluctuating quintessence field
  modifies the potential.  Since the potential involves in principle
  arbitrary powers of $\Phi$, we depict $V^{\prime\prime}$ as multiple
  external lines. }
\label{fig::pure}
\end{center}
\end{figure}

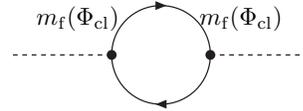
\begin{figure}[!t]
\begin{center}
\scalebox{0.93}[0.93]{
\begin{picture}(130,70)
\SetOffset(-97,-25)
\DashLine(100,60)(140,60){2}
\Vertex(140,60){2}
\ArrowArc(160,60)(-20,0,180)
\ArrowArc(160,60)(-20,180,0)
\Vertex(180,60){2}
\DashLine(180,60)(220,60){2}
\Text(110,75)[l]{{$\mf$}}
\Text(210,75)[r]{{$\mf$}}
\end{picture}}
\caption{Correction to the quintessence potential due to fermion fluctuations.
Fermion lines are solid, quintessence lines dashed.
Shown is the case where $m_{\rm f}(\Phi)$ gives a Yukawa coupling, i.e.
$c(\Phi) = \beta \Phi$, corresponding to one quintessence line. Of course,
for more complicated $m_{\rm f}(\Phi)$ such as  $\mf = m^0_{\rm f}\exp(-\beta \pmpc)$,
several external lines as in figure \ref{fig::pure}
would appear.
} \label{fig::fermion}
\end{center}
\end{figure}

By means of a saddle point expansion \cite{peskin}, we arrive at the
effective action $\Gamma[\pmpc]$ to one loop order of the quintessence
field.  The equation governing the dynamics of the quintessence field
is then determined by $\delta \Gamma[\pmpc] _{| \pmpc= \pmpc^\star} =
0$. When estimating the magnitude of the loop corrections, we will
assume that $\pmpc^\star$ is close to the solution of the classical
field equations: $\delta S = 0$.  Evaluating $\Gamma$ for constant fields,
we can factor out the space-time volume
$U$ from $\Gamma=UV$. This gives the effective potential
\begin{equation}\label{final}
\vv{\eloop}{}(\pmpc) = V(\pmpc) + \frac{\Lambda^2}{32\pi^2} V^{\prime\prime}(\pmpc) - \frac{\lferm^2 }{8\pi^2} \mfsq.
\end{equation}
Here, primes denote derivatives with respect to $\Phi$; $\Phi_{\rm
  cl}$ is the classical field value and $\Lambda$ and $\lferm$ are the
ultra violet cutoffs of scalar and fermion fluctuations. 
The last term in Equation \eqref{final} accounts for the fermionic
loop corrections as shown in figure \ref{fig::fermion}.
The second term
in Equation \eqref{final}, is the leading order scalar loop,
depicted in figure \ref{fig::pure1}.
We neglect graphs of the order  $(V^{\prime\prime}_{| \rm cl})^2$ and higher
like the one in figure \ref{fig::pure2}, because $V$ and its derivatives
are of the order $10^{-120}$ (see section \ref{sec::coupling}).  We have
also ignored $\Phi$-independent contributions, as these will not
influence the quintessence dynamics.

However, the $\Phi$-independent contributions add up to a cosmological
constant of the order $\Lambda^4 \approx {\mathcal{O}}(\mplank^4)$. This
is the old cosmological constant problem, common to most field
theories.  We hope that some symmetry or a more
fundamental theory will force it to vanish.  The same symmetries or
theories could equally well remove the loop contribution by
some cancelling mechanism. After all, this mechanism must be there,
for the observed cosmological constant is far less than the naively
calculated ${\mathcal{O}}(\mplank^4)$. Unfortunately, SUSY is broken
too badly to be this symmetry \cite{Kolda:1999wq}.

In addition, none of the potentials under investigation can be renormalized in the
strict sense. However, as we will see, terms preventing renormalization
may in some cases be absent to leading order in $V^{\prime\prime}_{| \rm cl}$.
As the mass of the quintessence
field is extremely small, one may for all practical purposes view these
specific potentials (such as the exponential potential) as
renormalizable.

There is also a loophole for all models that will be ruled out in
the following: The potential used in a given model could be the
full effective potential including all quantum fluctuations, down
to macroscopic scales. For coupled quintessence
models, this elegant argument is rather
problematic and the loophole shrinks to a point (see section
\ref{sec::coupling}).

In the following, we apply Equation \eqref{final} to various
quintessence models in order to check their stability against one
loop corrections. We do this separately
for coupled and uncoupled models. We use units in which $\mplank =
1$. For clarity, we restore it when appropriate.

\section{Uncoupled quintessence}\label{uncoupled}
Here, we are going to discuss inverse power law, pure and modified exponential,
and cosine-type potentials.
\subsection{Inverse power law and exponential potentials}
Inverse power laws \cite{Peebles:1988ek,Ratra:1988rm},
exponential potentials \cite{Wetterich:1988fm,Ferreira:1998hj} and mixtures of
both \cite{Brax:1999gp} can be treated by considering   the potential
$V = A \pmp^{-\alpha} \exp( -\lambda \pmp^\gamma)$ \cite{Corasaniti:2002mf}.
Limiting cases include inverse power laws,
exponentials, and  SUGRA inspired
models. Deriving twice with respect to $\Phi$, we find
\begin{multline}\label{zweimal}
V^{\prime\prime} =A \pmp^{-\alpha} \exp(- \lambda \pmp^\gamma) \Bigl \{
\alpha(\alpha+1) \Phi^{-2} + 2 \alpha  \lambda \gamma  \pmp^{\gamma-2}  \\
+ \lambda^2 \gamma^2 \pmp^{2\gamma -2} - \lambda \gamma(\gamma -1) \pmp^{\gamma-2}
\Bigr \}.
\end{multline}
\subsubsection{Inverse power laws}
For inverse power laws, we set $\gamma=\lambda =0$. This gives the
classical potential \mbox{$\vv{cl}{ipl} = A \pmpc^{-\alpha}$} and by means of
Equation \eqref{final} the loop corrected potential
\begin{equation}\label{equ::ipl}
\vv{\eloop}{ipl} = \vv{cl}{ipl}  \left(1 +   \frac{1}{32\pi^2}  \Lambda^2\alpha(\alpha+1) \pmpc^{-2}\right).
\end{equation}
The potential is form stable if $\frac{1}{32\pi^2} \Lambda^2\alpha(\alpha+1) \Phi^{-2} \ll 1$,
which today is satisfied, as $\Phi \approx \mplank$ \cite{Brax:1999gp}.

However, if the field is on
its attractor today, then $\Phi \propto (1+z)^{-3/(\alpha +2)}$, where $z$ is
the redshift \cite{Brax:1999gp}.
Using this, we have for $z \gg 1$
 \begin{equation}
\vv{\eloop}{ipl} \approx \vv{cl}{ipl}  \left(1 +   \frac{1}{32\pi^2} \Lambda^2\alpha(\alpha+1)  z^{6/(\alpha +2)}\right).
\end{equation}
Thus, the cutoff needs to satisfy
$\Lambda^2 \ll \frac{32\pi^2}{\alpha(\alpha+1)} \times z^{-6/(\alpha +2)}$.
Cosmologically viable inverse power law potentials seem to be
restricted to $\alpha < 2$ \cite{Balbi:2001kj,Doran:2001ty}. Using
$\alpha =1$ and $z\approx 10^4$ for definiteness, the bound becomes
$\Lambda^2 \ll  10^{-6}$.

So, at equality (and even worse before that epoch), the cutoff needs to
be well below $10^{12}$ Gev, if classical calculations are meant to be
valid. In \cite{Brax:2000yv} it is argued that for inverse power laws,
the quintessence content in the early universe is negligible and hence
the fluctuation corrections are important only at an epoch where
quintessence is subdominant. As the loop corrections introduce only
higher negative powers in the field, it is hoped that, even though one
does not know the detailed dynamics, the field will nevertheless roll
down its potential (which at that time is supposed to be much steeper)
and by the time it is is cosmologically relevant, the classical
treatment is once again valid.  Having no means of calculating the
true effective potential for the inverse power law in the early
universe, this view is certainly appealing.

\subsubsection{Pure exponential potentials}
The pure exponential potential is special because its derivatives are multiples of itself.
The classical potential (with $\alpha=0,\ \gamma=1$)
is $\vv{cl}{ep} =  A\exp( - \lambda \pmpc)$ and
to one loop order
\begin{equation}
\vv{\eloop}{ep} =  \vv{cl}{ep} \left \{ 1 +  \frac{1}{32\pi^2}  \Lambda^2 \lambda^2 \right\}.
\end{equation}
It is easy to see that a rescaling of $A \to A  / \left( 1 +   \frac{1}{32\pi^2}  \Lambda^2 \lambda^2 \right)$
absorbs the loop correction, leading to a stable potential up to order $V_{\rm cl}^{\prime\prime}$.
Working to next to leading order, i.e. restoring terms of order $(V_{\rm cl}^{\prime\prime})^2$,
we get
\[
\vv{\eloop,\, n.l.}{ep} =  \frac{1}{32 \pi^2}  \left(V_{\rm cl}^{\prime\prime}\right)^2  \ln \left(\frac{V_{\rm cl}^{\prime\prime}}{\Lambda^2} \right).
\]
It is this term which in four dimensions spoils strict renormalizability.

\subsection{Nambu-Goldstone cosine potentials}
Cosine type potentials resulting from a quintessence axion
were introduced in \cite{Kim:1999kx,Nomura:2000yk} and their implications for
the CMB have been studied in \cite{Kawasaki:2001bq}. They take on the
classical potential
$\vv{cl}{ng} = A \left [1 - \cos\left(\pmpc/ f_Q \right) \right]$
and including loop corrections
\[
\vv{\eloop}{as} = A \left[1 - \left\{ 1 - \frac{1}{32\pi^2}  \frac{ \Lambda^2}{f_Q^2}  \right \}
\cos\left(\frac{\Phi_{\rm cl}}{f_Q}\right)\right].
\]
Upon a redefinition $ A \to A /     \left\{ 1 - \frac{1}{32\pi^2}  \frac{ \Lambda^2}{f_Q^2}  \right \} $
and, recalling that the loop correction is only defined up to a constant, one arrives at the
same functional form as the classical potential.

\subsection{Modified exponentials}
In the model proposed by Albrecht and Skordis \cite{Albrecht:2000rm}, the classical potential is
$\vv{cl}{as} = V_p \exp ( - \lambda \pmpc)$, where $V_p$ is a
polynomial in the field. To one loop order, this leads to
\begin{equation} \label{ase}
\vv{\eloop}{as} =  \vv{cl}{as} \left\{1 + \frac{1}{32 \pi^2 } \Lambda^2 \left (
\frac{V_p^{\prime\prime}}{V_p} - 2 \lambda \frac{V_p^{\prime}}{V_p} + \lambda^2
 \right) \right\}.
\end{equation}
Let us for definiteness discuss the example given in \cite{Albrecht:2000rm},
where the authors chose $V_p(\Phi) = (\pmp - B)^2 + C$. With this choice,
we have
\begin{multline}
\vv{\eloop}{as,\,exmpl} =  \vv{cl}{as,\,exmpl}   \bigg\{1 + \frac{1}{32 \pi^2 } \Lambda^2
\Big (\frac{1}{V_p} \big [
2  \\
- 4\lambda( \pmpc - B) \big ] + \lambda^2
 \Big) \bigg\}.
\end{multline}
Now consider field values  close to the minimum of $V_p$, i.e., let the absolute value of
$\xi \equiv \pmpc - B$ be small compared to $\sqrt{C}$. Then
\begin{multline}
\vv{\eloop}{as,\,exmpl}  =   \vv{cl}{as,\,exmpl} \left\{1 + \frac{\Lambda^2 }{32 \pi^2 }\left (
\frac{2 - 4\lambda \xi  }{C + \xi^2}  + \lambda^2
 \right) \right\},
\end{multline}
and to leading order in $\xi$
\begin{equation}
\vv{\eloop}{as,\,exmpl}  \approx  \vv{cl}{as,\,exmpl}   \left\{1 +  \frac{\Lambda^2 }{32 \pi^2 } \left (\frac{1}{C} \left[
2 - 4\lambda \xi \right ] + \lambda^2
 \right) \right\}.
\end{equation}
Now consider, as was the case in the example given in \cite{Albrecht:2000rm}, $C=0.01$ for
definiteness. If we assume a cutoff $\Lambda$ and a Plank mass of approximately the same order, we get
\begin{equation}\label{asc}
\vv{\eloop}{as,\,exmpl}  \approx  \vv{cl}{as,\,exmpl}     \left\{1 + \frac{1}{32 \pi^2 } \left(100 \left[
2 - 4\lambda \xi \right ] + \lambda^2
 \right) \right\}.
\end{equation}
The $\xi$ (and hence $\pmpc$) dependent contribution in the curly brackets of Equation \eqref{asc}
is $-25/(2 \pi^2) \lambda \xi $ which for the value $\lambda =8$ chosen in the example gives
$-200/(2 \pi^2) \xi \approx -10 \xi$.

\begin{figure}
\begin{center}
\includegraphics[scale=0.35,angle=-90]{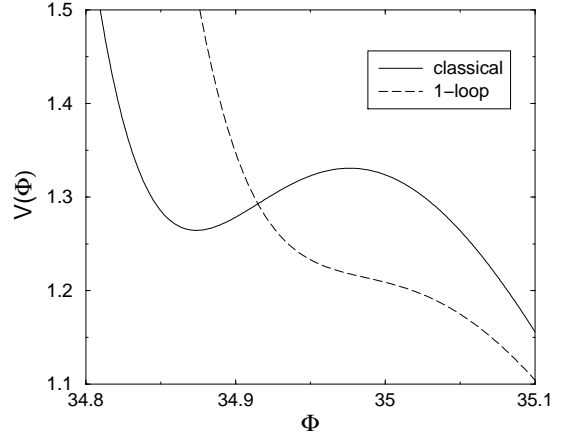}
\caption{Classical and 1-loop corrected potential [in $10^{-123} \mplank^4$] for
$\vv{cl}{as} = \left [ (\pmp - B)^2 + C \right] \exp ( - \lambda \pmpc)$ with
$B = 34.8,\ C=0.013,\ \Lambda = 1.2$. The classical potential has
a local minimum, which is absent for the loop corrected one. This is
a hand-picked example and in most cases, the bump will not vanish but
move and change its form.}\label{fig::albrecht}
\end{center}
\end{figure}

If we now look at the behaviour of the loop correction as a
function of $\pmpc$ and hence $\xi$ in the vicinity of the
minimum of this example polynomial, we see that for, e.g., $\xi =
0.01$, the one loop contribution dominates the classical potential
giving rise to a linear term in $\pmpc$
unaccounted for in the classical treatment. For many values of the
parameters $B$ and $C$, this just changes the form and location of
the bump in the potential.  In principle, however the loop
correction can remove the local minimum altogether (see figure
\ref{fig::albrecht}).

Needless to say that this finding depends crucially on the cutoff. If
it is chosen small enough, the conclusion is circumvented. In addition,
only the specific choice of $V_p$ above has been shown to be potentially
unstable. The space of polynomials is certainly large enough to provide
numerous stable potentials of the Albrecht and Skordis form.

\section{Coupled quintessence}\label{sec::coupling}
Various models featuring a coupling
of quintessence to some form of dark matter have been
proposed
 \cite{Wetterich:1995bg,Amendola:2000er,Bertolami:2000dp,Tocchini-Valentini:2001ty,Bean:2001zm}.
From the action Equation
\eqref{equ::action}, we see that the mass of the fermions could be $\Phi$
dependent: $m_{\rm f} = \mf$. Two possible realization of this mass
dependence are, for instance, $m_{\rm f} = m^0_{\rm f} \exp(-\beta
\pmpc)$ and $m_{\rm f} = m^0_{\rm f} + c(\Phi_{\rm cl})$, where in the second
case, we may have a large field independent part together with small
couplings to quintessence.\footnote{The constant $\mfn$ is \emph{not} the fermion mass today, which
would rather be $m_{\rm today} = m_{\rm f}(\pmpc({\rm today}))$.}
 For the model discussed in
\cite{Amendola:2000er}, the coupling is of the first form, whereas in \cite{Bean:2001zm},
the coupling is realized by multiplying the cold dark
matter Lagrangian by a factor $f(\Phi$).
This factor is usually taken of the form $f(\Phi) = 1 + \alpha(\Phi - \Phi_0)^\beta$.
Hence, the coupling is $m_{\rm f}(\Phi) = f(\Phi)\, \mfn$, if we assume that
dark matter is fermionic. If it were bosonic, the following arguments would be similar.

We will first discuss general bounds
on the coupling and in a second step check whether these bounds
are broken via an effective gravitational coupling.

\subsection{General bounds on a coupling}
We will discuss only the new effects coming from the coupling and
set
\begin{equation}
\vv{\eloop}{} = \vv{cl}{}  - \Delta V,
\end{equation}
where
$\Delta V = \lferm^2 \mfsq / \left(8\pi^2\right)$.
If we assume that the potential energy of the  quintessence field
constitutes a considerable part
of the energy density of the universe today, i.e. $\rho_{\rm q} \sim \rho_{\rm critical}$,
we see from the Friedmann equation
\begin{equation}
3 H^2 = \rho_{\rm critical},
\end{equation}
that
$\vv{cl}{} \sim  H^2$.
With today's Hubble parameter $H = 8.9 \times 10^{-61}\,h$ ($h = 0.5 \dots 0.9$),
we have
\begin{equation}\label{equ::vclassic}
V_{\rm cl} \sim 7.9 \times  10^{-121} \,h^2.
\end{equation}
The ratio of the `correction' to the classical potential is
\begin{equation}\label{overwhelming}
\frac{\Delta V}{\vv{cl}{}} = \frac{1}{8\pi^2} \frac{\lferm^2  \mfsq }{\vv{cl}{}}.
\end{equation}

Let us first consider the case that all of the fermion mass is
field dependent, i.e., we consider cases like
$m_{\rm f} = m^0_{\rm f} \exp(-\beta\pmpc)$.
As an example, we choose
a fermion cutoff at the GUT scale $\lferm =10^{-3}$,
and a fermion mass, $\mf$
of the order of $100\, \gev= 10^{-16} \mplank$. Then
Equation \eqref{overwhelming}
gives the overwhelmingly large ratio
\begin{equation}\label{equ::gross}
\frac{\Delta V}{\vv{cl}{}} \approx 10^{80}.
\end{equation}
Thus, the classical potential is negligible relative to the
correction induced by the
fermion fluctuations.

Having made this estimate, it is clear that the fermion loop
corrections are harmless only, if the square of the coupling takes on
\emph{exactly} the same form as the classical potential itself. If,
for example, we have an exponential potential
$\vv{cl}{} = A\exp(-\lambda \pmpc)$
together with a coupling $\mf = m^0_{\rm f}\exp(-\beta \pmpc)$, then
this coupling can only be tolerated, if $2\beta  = \lambda$.\footnote{Of course, a sufficiently small $\beta$ will lead to a more
or less constant contribution, where $\mf \approx  m^0_{\rm f} -\beta \pmpc$.}
Taken at face value, this finding restricts models with these types of coupling.
It is however interesting to note that for exponential coupling, the
case $2\beta = \lambda$ is not ruled out 
by cosmological \mbox{observations \cite{Tocchini-Valentini:2001ty}.}

Turning to the possibility of a fermion mass that consists of
a field independent part and a coupling, i.e.,
$m_{\rm f} = m^0_{\rm f} + c(\Phi_{\rm cl})$,
Equation \eqref{overwhelming} becomes
\begin{equation}\label{o2}
\frac{\Delta V}{\vv{cl}{}} = \frac{1}{8\pi^2} \frac{ \lferm^2  \left[ 2m^0_{\rm f} c(\Phi_{\rm cl}) +c(\Phi_{\rm cl})^2\right] }{\vv{cl} \,},
\end{equation}
where we have ignored a quintessence field independent
contribution proportional to $(m^0_{\rm f} )^2$. Assuming
$c(\Phi_{\rm cl}) \ll m^0_{\rm f} $, and demanding that the
loop corrections should be small compared to the classical potential,
Equation \eqref{o2} yields the bound
\begin{equation}\label{equ::cbound}
c(\Phi_{\rm cl})  \ll \frac{4 \pi^2 \, \vv{cl} \, }{ \lferm^2  m^0_{\rm f}}.
\end{equation}
If, as above, we assume $\lferm = 10^{-3} \mplank$, $m^0_{\rm f} = 10^{-16}\mplank$
and $\vv{cl}{}$ from Equation \eqref{equ::vclassic},
this gives
\begin{equation}
c(\Phi_{\rm cl})  \ll 3 \times 10^{-97},
\end{equation}
in units of the Planck mass. Once again, the bound from Equation \eqref{equ::cbound}
applies only  if the functional form of the loop correction differs from the
classical potential.
Assuming a Yukawa-type coupling $c(\pmpc) = \beta \pmpc$
and field values of at least the order of the Plank mass,
we get $\beta \ll 10^{-97}$.\

For the coupling   $c(\Phi)=\mfn\alpha(\Phi-\Phi_0)^{\beta}$ with the values
$\alpha=50,\,\beta=8,\,\Phi_0=32.5 $ given in \cite{Bean:2001zm}, $c(\Phi)$
is usually larger than $\mfn$. Therefore we take $\mf \approx c(\pmpc)$.
With  $\mf=10^{-16}$ as before, we get the same result as \mbox{in \eqref{equ::gross}.}

The coupled models share one property: the loop contribution from the
coupling is by far larger than the classical potential. At first sight,
the golden way out of this seems to be to view the potential as already
effective: all fluctuations would be included from the start.
However, there is no particular reason, why \emph{any} coupling
of quintessence to dark matter should produce just exactly \emph{the}
effective potential used in a particular model: there is a relation between
the coupling and the effective potential generated.
Put another way,
\emph{if} the effective potential is of an elegant form and we have
a given coupling, then it seems unlikely that the \emph{classical}
potential could itself be elegant or natural.

\subsection{Effective gravitational fermion quintessence coupling}\label{gravcoupling}
\begin{figure}[!t]
\begin{center}
\subfigure[]{\scalebox{0.93}[0.93]{
\begin{picture}(110,100) \label{fig::gravi1}
\SetOffset(-108,-15)
\Text(160,110)[]{{$V(\pmpc)$}}
\ArrowLine(110,20)(130,40)
\Vertex(130,40){2}
\Photon(130,40)(160,70){2.5}{4.5}
\Photon(160,70)(190,40){2.5}{4.5}
\Vertex(160,70){2}
\Vertex(190,40){2}
\ArrowLine(130,40)(190,40)
\ArrowLine(190,40)(210,20)
\DashLine(160,70)(175,95.98){2}
\DashLine(160,70)(167.97,98.98){2}
\DashLine(160,70)(152.23,98.98){2}
\DashLine(160,70)(145,95.98){2}
\Text(161.2,92.5)[]{\tiny{$\cdots$}}
\end{picture}}}
\subfigure[]{\scalebox{0.93}[0.93]{
\begin{picture}(110,100) \label{fig::gravi2}
\SetOffset(-108,-15)
\Text(160,110)[]{{$V(\pmpc)$}}
\ArrowLine(120,20)(160,40)
\Vertex(160,40){2}
\Vertex(160,70){2}
\PhotonArc(160,55)(15,-90,90){-2.5}{5.5}
\PhotonArc(160,55)(15,90,270){-2.5}{5.5}
\ArrowLine(160,40)(200,20)
\DashLine(160,70)(175,95.98){2}
\DashLine(160,70)(167.97,98.98){2}
\DashLine(160,70)(152.23,98.98){2}
\DashLine(160,70)(145,95.98){2}
\Text(161.2,92.5)[]{\tiny{$\cdots$}}
\end{picture}}}
\caption{Effective fermion-quintessence coupling via graviton exchange.
The fermions (solid lines) emit gravitons (wiggled lines) which are
caught by the quintessence field (dashed lines). As the graphs involve
couplings of the gravitons to the classical quintessence potential, the
generated coupling is proportional to the classical  potential. Since the potential
involves arbitrary powers of $\Phi$, we depict it as several $\Phi$-lines.
A Yukawa type coupling, corresponding to just one line, is then generated
 by power expanding  $V(\Phi) = V(\pmpc){} + V_{|\textrm{cl}}^{\prime}\, (\Phi - \pmpc)$
 in the fluctuating field.
} \label{fig::gravi}
\end{center}
\end{figure}
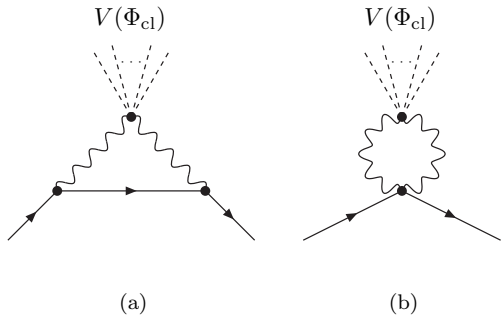
The bound in Equation \eqref{equ::cbound} is so severe that the question
arises whether gravitational coupling between fermions and the
quintessence field violates it.
To give an estimate, we calculate\footnote{Unfortunately, the field-dependent 
propagator matrix is non-diagonal ($\pmpc \neq 0$ usually). This is a subtle point. 
We split the full propagator into a field-independent part $P$ and a field-dependent part $F$. The
logarithm in ${\rm STr} \log (P + F)$ is then expanded in powers of $F$. For the  Weyl-Frame 
calculation in Section \ref{sec::weyl}  this is not longer possible, as the 
graviton-graviton propagator involves the field $\chi^2$ and thus the field-independent
part $P$ is non-invertible. For simplicity, we ignored the gravity part in the Weyl-Frame 
calculation (including the coupling of gravitons to $\chi$).} 
two simple processes depicted in figure \ref{fig::gravi}.
We evaluate the diagrams for vanishing external momenta. This is consistent
with our derivation of the fermion loop correction Equation \eqref{final}, in which we
have assumed momentum independent couplings.
The effective coupling due to the graviton exchange contributes
to the fermion mass, which becomes $\pmpc$ dependent. We assume that
this coupling is small compared to the fermion mass and write
$\mf = \mfn  + c(\pmpc)$.

\begin{figure}[!t]
\begin{center}
\scalebox{0.93}[0.93]{
\begin{picture}(130,80)
\SetOffset(-100,-23) \Text(110,90)[l]{$V(\pmpc)$}
\DashLine(150,60)(115.36,40){2}
\DashLine(150,60)(111.36,49.65){2}
\DashLine(150,60)(115.36,80){2}
\DashLine(150,60)(111.36,70.35){2}
\Text(120,63)[]{$\vdots$}
\Photon(150,60)(175.86,45.86){-2.5}{4.5}
\Photon(150,60)(175.86,74.14){2.5}{4.5}
\Vertex(175.86,45.86){2}
\Vertex(175.86,74.14){2}
\ArrowArc(190,60)(-20,45,180)
\ArrowArc(190,60)(-20,180,315)
\ArrowLine(175.86,45.86)(175.86,74.14)
\Vertex(210,60){2}
\Line(214,64)(206,56)
\Line(214,56)(206,64)
\end{picture}}
\caption{Fermion loop contribution to the quintessence potential involving the effective coupling figure \ref{fig::gravi1}.
The cross in the fermion line depicts the field independent fermion mass $\mfn$.}
\label{fig::effective}
\end{center}
\end{figure}
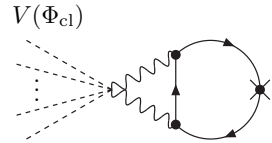

From the first diagram, figure
\ref{fig::gravi1} we get (see the \mbox{Appendix \ref{app::weyl}):}
\begin{multline}\label{equ::gravi2}
c(\pmpc)=
\frac{1}{8\pi^2}m_{\textrm{f}}^0V(\pmpc)
\times \Bigg[
\ln\left(\frac{\Lambda^2}{\Lambda^2 + \mfnsq}\right)\\
-\ln\left(\frac{\lambda^2}{\lambda^2 + \mfnsq}\right) \Bigg],
\end{multline}
whereas \ref{fig::gravi2} gives
\begin{equation}\label{equ::gravi1}
c(\pmpc)=
\frac{5}{8\pi^2}m_{\textrm{f}}^0V(\pmpc)\ln\left(\frac{\Lambda}{\lambda}\right).
\end{equation}
Here, we have introduced infrared and ultraviolet cutoffs $\lambda$
and $\Lambda$ for the graviton
momenta.  We assume $\Lambda$ to be of the order $\mplank$
and $\lambda$ about the  inverse size of the horizon. Since
the results depend only logarithmically on the cutoffs, this choice
is not critical, and in addition $\ln(\mplank / H) \approx 140$, which is small.
From Equation (\ref{o2}, \ref{equ::gravi1}, \ref{equ::gravi2}),
we see that, in leading order, the change in the quintessence
potential due to this effective fermion coupling would be proportional
to $V(\pmpc)$ and could hence be absorbed upon
redefining the pre-factor of the potential (see also figure \ref{fig::effective}).
In next to leading order,
the contribution is proportional to $V(\pmpc)^2$, which is negligible.

From the Appendix \ref{app::weyl}, in which we present the calculation in more detail,
it is clear that there are processes where the vertices are more complicated.
However, to this order all diagrams are proportional to $V(\pmpc)$. Thus,
they can be absorbed just like the two processes presented above.

\section{Weyl transformed fields}\label{sec::weyl}
So far, we have assumed a constant Planck mass together with a
field independent cutoff. We could, however, assume that the Planck
mass is not constant, but rather
given by the expectation value of a scalar field $\chi$. We will
call the frame resulting from this Weyl scaling the Weyl frame,
as opposed to the frame with a constant Plank mass which we will call
the Einstein frame. From the classical point of view, both frames are
equivalent.  On calculating quantum corrections, we have to
evaluate a functional integral. Usually, the functional measure in
the Einstein frame is set to unity. In principle, the variable
change associated with the Weyl scaling leads to a non-trivial
Jacobian and therefore a different functional measure. Taking
the position that the Weyl frame is fundamental, this measure
could equally well be set to unity in the Weyl frame.
Therefore, it is a priori unclear whether the loop corrected
potential in the Weyl frame, when transformed back into the
Einstein frame, will be the same as the one from Equation
\eqref{final}.

As the cutoff in the Einstein frame is a constant mass scale and
hence proportional to the Plank mass, it seems natural to assume
that the cutoff in the Weyl frame is proportional to $\chi$. We restrict
our discussion to this case. For other choices of the $\chi$-dependence 
of the cutoff, the results may differ.

The Weyl transformation is achieved by
scaling the metric, the curvature scalar,
all fields, and the tetrad by appropriate powers of $\chi / \mplank$ (see table \ref{tab::weyl})
\cite{Wetterich:1988fm,Wetterich:1995bg}:
\begin{multline}\label{equ::weyl}
\tilde S =  \int d^4x \sqrt{\tilde g}
\Bigg [ \chi^2 \tilde R + \frac{z}{2}\partial_\mu\chi\partial^\mu\chi
+ W(\chi) \\
+ \tilde{\bar\Psi}\left(i\,\tilde\gamma^\mu(x) \nabla_\mu + \chi \frac{\mf}{\mplank} \gamma^5 - \frac{3}{2}i\tilde\gamma^\mu(x)\ln \chi_{,\mu}\right)\tilde\Psi \Bigg ],
\end{multline}
where $\Phi = (12+z)^{1/2}\mplank\ln(\chi/\mplank)$ and
\begin{equation}
 W(\chi) \equiv \left(\frac{\chi}{\mplank}\right)^4 V(\Phi(\chi)).
\end{equation}
The term proportional to $\ln \chi_{,\mu}$ in Equation \eqref{equ::weyl}
is somewhat inconvenient. Adopting the position that the Weyl frame is
fundamental, this term is unnatural. Instead, one could formulate the
theory with canonical couplings for the fermions.
Dropping this term,
\begin{multline}\label{equ::weylcan}
\tilde S_{\textrm{can.}} = \int d^4x \sqrt{\tilde g}
\Bigg [ \chi^2 \tilde R + \frac{z}{2}\partial_\mu\chi\partial^\mu\chi
+W(\chi)\\
+ \tilde{\bar\Psi}\left(i\,\tilde\gamma^\mu(x) \nabla_\mu + \frac{\chi}{\mplank} m_{\rm f}(\Phi(\chi))\, \gamma^5 \right)\tilde\Psi \Bigg ],
\end{multline}
we observe by going back to the usual action $\tilde S_{\textrm{can.}} \to S$,
 \begin{multline}\label{action2}
S  = \int d^4x\   \sqrt{g} \Bigg[\frac{1}{2} \partial_{\mu} \Phi(x) \partial^{\mu} \Phi(x) + V(\Phi(x)) \\
+ \bar\Psi(x)\left(i \slad{\nabla} + \gamma^5 m_{\rm f}(\Phi)  + \frac{3}{2\mplank}i\,\gamma^\mu(x) \phi_{,\mu}\right )  \Psi(x) \Bigg],
\end{multline}
that the canonical form of the action in the Weyl frame gives
rise to a derivative coupling of the quintessence field to the fermions
in the Einstein frame,
which we can safely ignore.\footnote{Actually, this coupling is non-renormalizable in the strict sense. Since the
theory is non-renormalizable anyway, this is not of great concern. In addition, if one believes
that the Weyl frame is fundamental, there is no need to go back to the Einstein
frame and hence no need to face this nuisance.}

\begin{table}
\begin{tabular}{r@{\ $\to$\ }l}
$g_{\mu\nu}$ & $(\chi / \mplank)^{2} \, \tilde g_{\mu\nu}$ \\[1.5ex]
$g^{\mu\nu}$ & $(\chi / \mplank)^{-2} \, \tilde g^{\mu\nu}$ \\[1.5ex]
$\sqrt{g}$ & $(\chi / \mplank)^{4} \, \sqrt{\tilde g}$ \\[1.5ex]
$R$ &  $\left(\chi/ \mplank \right)^{-2} \left( \tilde R - 6 \tilde g^{\mu\nu} \sigma_{\tilde; \mu \nu} - 6  \tilde g^{\mu\nu} \sigma_{,\mu} \sigma_{,\nu}\right)$\\[1.5ex]
$e^{\mu}_a(x)$ & $(\chi / \mplank)^{-1} \, \tilde e_a^\mu(x)$ \\[1.5ex]
$\Psi$ & $(\chi / \mplank)^{-3/2}\, \tilde\Psi$ \\[1.5ex]
\end{tabular}
\caption{Weyl scaling of various quantities. The transformation of the
curvature scalar $R$ follows from the scaling of the metric. This
scaling, in turn,
originates from the condition that instead of the Plank mass squared
multiplying $R$ in the action in the Einstein frame, a factor $\chi^2$ should appear.
Here, we have set $\sigma = \ln(\chi / \mplank)$.
}\label{tab::weyl}
\end{table}

Working with Equation \eqref{equ::weylcan}, we get the loop correction
in the Weyl frame by replacing $V \to W$ and $\Phi \to \chi$ in Equation \eqref{final}.
In addition, the constant cutoffs $\Lambda$ and $\lferm$ are replaced by ${ const} \cdot \chi$:
\begin{multline}\label{equ::weylloop}
\ww{\eloop}{} = W(\chi) + \frac{\left({\mathcal{C}}\chi \right) ^2 }{32\pi^2\, z^2} W^{\prime\prime}(\chi) \\- \frac{\left({\mathcal{C}}_{\rm f} \chi\right)^2}{8\pi^2}
\left [\frac{\chi}{\mplank} m_{\rm f} (\chi) \right ] ^2.
\end{multline}
Transforming $\ww{\eloop}{}$ back into the Einstein frame, the
potential $V$ is modified by
\begin{multline}\label{equ::back}
\vv{1-loop}{} = V(\pmpc) +  \frac{\left({\mathcal{C}}_{\rm f}\mplank\right)^2}{8 \pi^2}\mfsq
+\frac{\left({\mathcal{C}} \mplank\right)^2}{32 \pi^2\,z^2} \\
\times \bigg [ 12 \frac{V(\pmpc)}{\mplank^2} +
7\sqrt{12+z} \frac{V^{\prime}(\pmpc)}{\mplank}
+ (12+z) V^{\prime\prime}(\pmpc) \bigg].
\end{multline}
As an example, lets calculate the correction to the pure exponential
potential $\vv{cl}{ep} =  A\exp( - \lambda \pmpc)$, once again
setting $\mplank = 1$. The Weyl frame potential is
\begin{equation}
W(\chi) = A \chi^4 \exp(-\lambda \pmpc(\chi))
= A \chi^{(4 - \lambda \sqrt{12 +z})}.
\end{equation}
Neglecting fermion fluctuations and choosing $z=1$,
\begin{equation}
W_{\rm 1-loop} = \left[ 1 + \frac{{\mathcal{C}}^2}{32 \pi^2\,z^2}
 (4 -  \lambda \sqrt{13}) (3 -  \lambda \sqrt{13})
 \right] W(\chi).
\end{equation}
Again (and not surprisingly) we can absorb the terms in the square brackets
in a redefinition of the pre-factor $A$. In the case of an inverse power
law, the term proportional to $V^{\prime}$ in Equation \eqref{equ::back}
leads to a slightly different contribution compared to Equation \eqref{equ::ipl}
(a term $\propto \pmpc^{-\alpha-1}$ arises).
For the modified exponential potentials the expressions corresponding
to $V^{\prime}$ in Equation \eqref{equ::back} make no structural difference.

\section{Conclusions}

We have calculated quantum  corrections to the classical
potentials of various quintessence models. In the late universe,
most potentials are stable with respect to the scalar quintessence
fluctuations.
The pure exponential  and Nambu-Goldstone type potentials
are form invariant up to order $V^{\prime\prime}$, yet terms
of order  $(V^{\prime\prime})^2$ prevent  them from being renormalizable in the strict sense.

For the modified exponential potential introduced by Albrecht and
Skordis, stability depends on the specific form of the polynomial
factor $V_p$ in the potential. In some cases the local minimum in the
potential can even be removed by the loop.

An explicit coupling of the quintessence field to fermions (or similarly to dark matter
bosons) seems to be severely restricted. The effective potential to one loop level would be
completely dominated by the contribution from the fermion fluctuations.
All models in the literature share this fate. One way around this conclusion
could be to view these potentials as already effective. They must, however, not only
be effective in the sense  of an effective quantum field theory originating as a low-energy limit of
an underlying theory, but also include all fluctuations from this effective QFT.
In this case, there is a strong connection between coupling and potential and it
is rather unlikely that the \emph{correct} pair can be guessed.

The bound on the coupling is so severe that for consistency, we
have calculated an effective coupling due to graviton exchange.
To lowest order in $V(\Phi)$, this coupling leads to  a fermion contribution
which can be absorbed by redefining the pre-factor of the potential.

To check that the results are not artefacts from the Einstein frame,
we switched to the Weyl frame. As the transition from $\Phi \, \to\, \chi$
involves a non-trivial Jacobian, the details of the results differ. However,
the basic results stay the same.

Surely, the one-loop calculation does not give the true effective potential.
Symmetries or more fundamental
theories that make the cosmological constant as small as it is, could
force loop contributions to cancel. In addition, the back reaction
of the changing effective potential on the fluctuations remains
unclear in the one loop calculation. A renormalization group
treatment would therefore be of great value. We leave this to
future work.

\acknowledgments
We would like to thank Gert Aarts, Luca Amendola, J\"urgen Berges, Matthew Lilley,
Volker Schatz, and Christof Wetterich for helpful discussions.

\begin{appendix}
\section{Coupling to gravitons}\label{app::weyl}
Fermions in general relativity are usually treated within
the tetrad formalism. The $\gamma$ matrices become
space-time dependent:
$\gamma^\mu(x) \equiv \gamma^a e_a^\mu(x)$.
Together with the spin connection $\omega$,
one uses
(see, e.g., \cite{zinn-justin,veltman}):
\begin{equation}\label{nabla}
\slad{\nabla}=e^{\mu}_{a}(x)\gamma^{a}\left(\partial_{\mu}+\frac{i}{4}
\sigma_{bc}\omega^{bc}_{\mu}\right).
\end{equation}
The action \eqref{equ::action} can then be expanded in small
fluctuations around flat space:  $g_{\mu\nu}=\delta_{\mu\nu}+h_{\mu\nu}/\mplank$.

Using the gauge fixing term
$-\frac{1}{2}( \partial^\nu h_{\mu \nu} - \frac{1}{2} \partial_\mu h_\nu^\nu)^2$
and expanding the action to second order in $h$, we find the propagator
\cite{veltman}:
\begin{equation}
P^{-1}_{\textrm{grav}}(k)=\frac{\delta_{\mu\alpha}\delta_{\nu\beta}
+\delta_{\mu\beta}\delta_{\nu\alpha}-\delta_{\mu\nu}\delta_{\alpha\beta}}{k^2}.
\end{equation}
The diagrams in figure \ref{fig::gravi} are generated by
the expansion of
$\sqrt{g}=1+\frac{1}{2}h^{\mu\mu}-\frac{1}{4}(h^{\mu\nu})^2 +\frac{1}{8}(h^{\mu\mu})^2$
multiplying the matter Lagrangian.
Additional (and more complicated) vertices originate from
the spin connection and the tetrad.

However, we do not consider external graviton lines, which would only
give corrections to the couplings and wave function renormalization of
the gravitons.  Therefore only internal gravitons appear.  In order to
contribute a quintessence dependent part to the fermion mass, the
gravitons starting from the fermion-graviton vertices (complicated as
they may be) have to touch quintessence-graviton vertices.  As these
quintessence vertices are proportional to $V(\pmpc)$, all diagrams to
lowest order in $V(\pmpc)$ will only produce mass contributions
proportional to $V(\pmpc)$.

Evaluating the diagrams in figure \ref{fig::gravi} for vanishing external momenta we
get \eqref{equ::gravi2} and \eqref{equ::gravi1}.
\end{appendix}

\end{document}